\documentclass[]{spie}  %>>> use for US letter paper
\usepackage{amsmath,amsfonts,amssymb}
\usepackage{graphicx}
\usepackage[colorlinks=true, allcolors=blue]{hyperref}

\title{Beam combination schemes and technologies for the Planet Formation Imager \footnote{Copyright 2016 Society of Photo-Optical Instrumentation Engineers. One print or electronic copy may be made for personal use only. Systematic reproduction and distribution, duplication of any material in this paper for a fee or for commercial purposes, or modification of the content of the paper are prohibited. DOI: http://dx.doi.org/10.1117/12.2232656}}

\author[a,b]{Stefano Minardi}
\author[c]{Sylvestre Lacour}
\author[d]{Jean-Philippe Berger}
\author[e]{Lucas Labadie}
\author[f]{Robert R. Thomson}
\author[g]{Chris Haniff}
\author[h]{Michael Ireland}
\affil[a]{Institute of Applied Physics, Friedrich Schiller University, Max-Wien-Platz 1, 07743 Jena, Germany}
\affil[b]{Leibnitz-Institut f\"ur Astrophysik Potsdam, An der Sternwarte 16, 14482 Potsdam, Germany}
\affil[c]{LESIA, CNRS/UMR-8109, Observatoire de Paris-Meudon, 5 place Jules Janssen, 92195  Meudon, France}
\affil[d]{European Southern Observatory, Karl-Schwarzschild-Str. 2, 85748 Garching bei M\"unchen, Germany}
\affil[e]{1st Institute of Physics, University of Cologne, Z\"ulpicher Str. 77, 50937 Cologne, Germany}
\affil[f]{Scottish Universities Physics Alliance (SUPA), Institute of Photonics and Quantum Sciences (IPaQS), Heriot Watt University, Riccarton, Edinburgh, EH14 4AS}
\affil[g]{Cavendish Laboratory, JJ Thomson Avenue, Cambridge CB3 0HE, United Kingdom}
\affil[h]{Research School of Astronomy \& Astrophysics, Mount Stromlo Observatory, Cotter Road, Weston Creek, ACT 2611, Australia}

\authorinfo{Further author information: (Send correspondence to S.M.)\\S.M.: E-mail: sminardi@aip.de, Telephone: +49 (0) 331 7499 687}

\begin{document} 
\maketitle

\begin{abstract}
The Planet Formation Imager initiative aims at developing the next generation large scale facility for imaging astronomical optical interferometry operating in the mid-infrared. Here we report on the progress of the Planet Formation Imager Technical Working Group on the beam-combination instruments. We will discuss various available options for the science and fringe-tracker beam combination instruments, ranging from direct imaging, to non-redundant fiber arrays, to integrated optics solutions. Besides considering basic characteristics of the schemes, we will investigate the maturity of the available technological platforms at near- and mid-infrared wavelengths.
\end{abstract}

% Include a list of keywords after the abstract 

\section{INTRODUCTION}
\label{sec:intro}  % \label{} allows reference to this section

The Planet Formation Imager (PFI) initiative\cite{Monnier:2014,mon16,ire16} is aiming at developing the next generation large scale facility for imaging astronomical optical interferometry 
at mid-infrared wavelength. The main scientific goal of the PFI facility will be the high-angular-resolution characterization of planet forming regions and young Jupiter-mass exoplanets 
in the neighborhood of our Sun\cite{Kraus:2014,kra16}. 
This science case justifies the choice of the operating optical bands (from L to Q), due to the favorable contrast between the central star and the 
low-mass companions achievable at mid-infrared wavelengths.
Due to the long operating wavelengths, maximal baselines in the order of 1 Km will be necessary to resolve the Hill sphere ({\it i.e.} the radius of the gravitational sphere of influence of a 
forming planet) of a Jupiter mass planet, estimated to be approximatly $\sim 2.5$ mas at the distance of the nearest star forming region (140 pc).

A conceptual technical study is currently underway to identify suitable key technologies for the realization of the PFI facility (see Ireland et al. 2016\cite{Ireland:2016} 
for an overview of the current status of the technical study).  
Here we report on the progress of the Technical Working Group (TWG) on the beam-combination instruments. The goal of this TWG is to identify eventually the architecture and 
underlying technology for the beam combination instruments needed for PFI.
Our working baseline scenario considers an array of 12 to 21, adaptive-optics-equipped telescopes and two alternative options for the science beam combination, namely a dispersed 
heterodyne\cite{Ireland:2014}(bands N and Q) or a dispersed homodyne scheme (L,M,N bands). In both cases, fringe tracking at near-infrared wavelengths will be necessary to access faint targets, thus requiring an additional beam-combiner  instrument, most probably operating in the near-infrared.

This paper will review the state-of-the-art of homodyne multi-telescope beam combiners encompassing bulk, fiber and integrated optical solutions (Section 2). 
In Section 3 we will use a simple numerical model to compare the intrinsic per-baseline sensitivity of three different beam combination architectures assuming a coherence retrieval 
algorithm based on the Visibility to Pixel Matrix formalism (V2PM\cite{Tatulli:2007}).
In Section 4 we will discuss the state-of-the-art of photonics technologies for the mid-infrared which could enable the manufacturing of the science beam-combiner for PFI. A 
discussion of the present technological challenges and a tentative roadmap for the development of a mid-infrared beam combination instrument measuring most of the available 
baselines at PFI are proposed in Section 5. Conclusions and recommendations are presented in Section 6.

\section{REVIEW OF HOMODYNE BEAM COMBINATION INSTRUMENTS}

In this Section, we will present existing interferometric beam combination solutions broadly classified by key technologies and/or combination concepts.
In particular, we will consider beam combiners based on bulk optics, optical fibers and integrated optics technologies, as well as discussing the concept of direct imaging.
Throughout the text, we will also adopt the conventional high-level classification of multi-telescope beam combiners according to the fringe and baseline encoding (see 
Le Bouquin et al. 2004\cite{LeBouquin:2004}). 
Fringe encoding can be temporal, spatial or matricial depending on whether the interference fringes are measured in temporal/spatial domain or recorded from the outputs 
of a phase-shifting interferometry set-up. Notice that spatial and matricial schemes can also be operated in temporal fringe scanning mode at the expense of a lower sensitivity.
The baseline encoding can be pairwise, partial or all-in-one. Pairwise combiners encode at each output fringes resulting from a single pair of telescopes. In partial encoding    
the baselines are divided among several partial beam combiners. All-in-one schemes multiplex fringes from all possible baselines at each individual output. 

\subsection{Bulk-optics combiners}

A general advantage of bulk optics beam combiners is that they can be designed to achieve high transmission with highly achromatic response.
Nonetheless they tend to be more voluminous than \textit{e.g.} fibered or integrated optics beam combiners, which potentially raises manufacturing and operation costs, due to the larger size of the cryo-vacuum vessel required to house the cold optics. 

A bulk optics pairwise combiner with traditional design (cascade of beam splitters/combiners) would almost certainly be prohibitively complex for $N_{\rm t}\ge$12 telescopes, 
as appears to be needed for PFI. However a highly simplified design for a white-light, pairwise multi-telescope combination measuring simultaneously all possible baselines has 
been proposed by Ribak et al. 2007\cite{Ribak:2007} and successfully tested in the laboratory for up to 6 channels. This design could be easily scaled up to a large number of 
apertures. However, the 2D spatial fringe patterns readout requires large detectors and the implementation of a high-resolution spectro-interferometric setup is not straightforward. 

\begin{figure}
\includegraphics[width=0.45\textwidth]{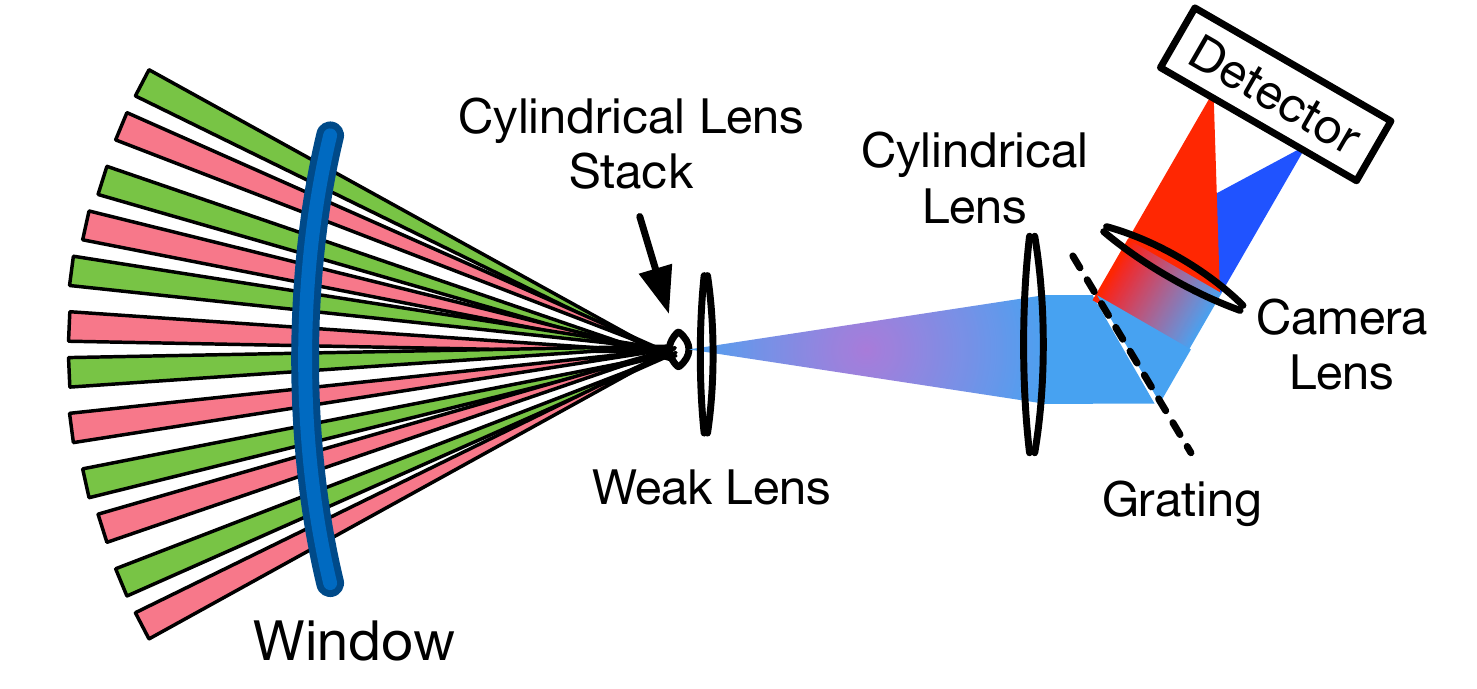}\quad
\includegraphics[width=0.45\textwidth]{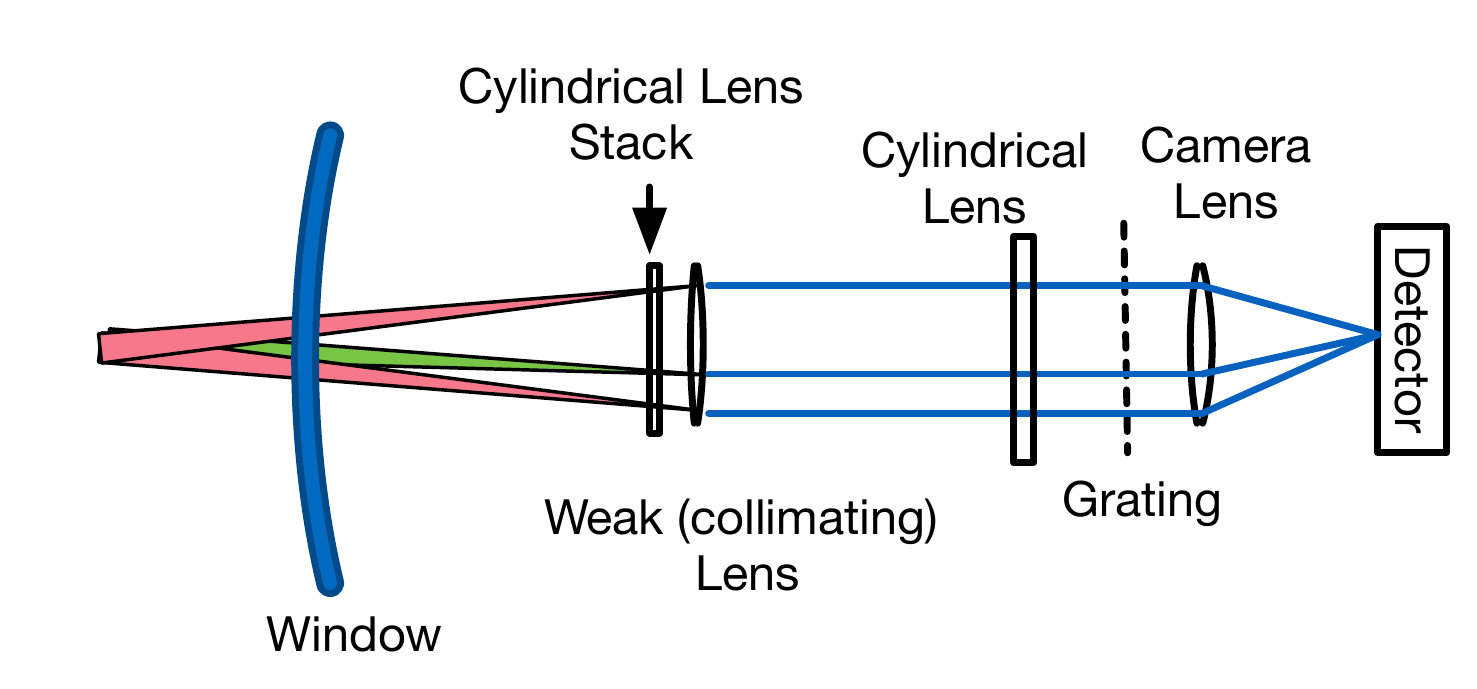}
\caption{Top down (Left) and side-on (Right) views of a 12-beam bulk-optics all-in-one combiner based 
on the non-redundant, multi-axial combination scheme. The full width from window to detector is $<
$1\,m. Left: Dispersion direction schematic, showing how a stack of offset cylindrical lenses could take 
beams from physically different locations outside a dewar, and co-align them. Right: In the fringe 
direction, the beam (with 3 of the 12 beams shown) directions are arranged on a non-redundant grid to 
extract the visibility of all possible baselines from the analysis of the interferogram. 
The two cylindrical components appear now as plane parallel glass blocks, and it is clear that all beams 
can originate from the same height outside the dewar. The element marked "Camera Lens" is almost 
certainly a multi-element camera, and may be a reflective camera.}
\label{figBulkNR}
\end{figure}

Existing or to-be-commissioned multi-telescope bulk optics beam combiners (\textit{e.g.} AMBER
\cite{AMBER} or MATISSE\cite{MATISSE}) are of the all-in-one type, based on an 
architecture where a linear non-redundant array of beams is combined in a multi-axial geometry and 
dispersed in the orthogonal direction\cite{Bedding:1994}. 
The multi-axial beam interference generates a spatial fringe pattern where the mutual coherence of 
each telescope pair is encoded in a unique spatial frequency, as a result of the 
non-redundant beam incidence angle selection. Multi-axial combiners necessarily requires cylindrical 
optics, with an anamorphic factor of order 
$N_{\rm baselines}=N_{\rm t}(N_{\rm t}-1)/2$ needed. 
A sketch of a 12-beam all-in-one multi-axial combiner is shown in Fig.~\ref{figBulkNR}. All-in-one multi-
axial combiners have the advantage of adding no additional optics per 
beam when increasing the number of beams. However the linear length of the non-redundant array 
scales faster than $N_{\rm t}^2$ (see following discussion on fibered beam 
combiners) and the diameter of the focusing optics required for fringe pattern generation could easily 
exceed 0.3-0.5 m for a PFI scenario. 
Because on-field experience has show that aberrations of the fringe imaging system substantially limit 
the performance of the multi-axial beam combiner, design and 
manufacturing of large aperture optical elements suitable for the beam combination task could be 
challenging.

Pupil plane, all-in-one multi-telescope beam combiners based on assemblies of multiple semi-reflecting 
mirrors have been used in the past to combine the beams at the 
COAST\cite{Haniff:2004} and NPOI\cite{Armstrong:1998} interferometers. This beam combiner is very 
stable due to the absence of moving parts and is operated in temporal scanning mode. 
Proposed upgrades of this design to measure simultaneously 
visibilities of up to 6 telescopes featured however footprints in the order of ~0.5 m$^2$ (Baron et al. 
2006\cite{Baron:2006}, Buscher et al. 2008\cite{Buscher:2008}), 
posing scaling constraints of the cold optics enclosure similar to the the multi-axial architecture in case 
a much larger number of telescopes are combined simultaneously.
However, instruments combining simultaneously a relatively small number of telescopes can be 
used as partial combiners for larger arrays by cycling the combined telescope multiplets with a fast 
reconfigurable switchyard\cite{Baron:2006}.

\subsection{Direct imaging instruments}
Conventional homodyne/heterodyne interferometric instruments measure the complex coherence function of
electromagnetic fields, from which images of the astronomical target can be retrieved.
The concept of pupil densification and the hyper-telescope \cite{labeyrie1996} can be exploited to 
design an instrument forming the convolution of the object and a relatively compact point spread function 
(PSF) on the detector, which can be interpreted as an image of the object.

The densification process defines how an input pupil will be remapped
into an output pupil that respects the overall telescope distribution
(the inner pupil central positions are homothetic to the outer ones) but
scales up the ratio between the output pupil diameters and the
distances between pupils. The result of the pupil densification is a reduction of the 
ratio between the field of view of the reconstructed image and the interferometric PSF. 
%Figure \ref{fig:densification} from \citet{patru2009} shows the comparison between a
%dilluted pupil distribution and a densified one. The maximum
%densification factor $\gamma_{\rm{max}}$ is defined by the ratio
%between the smallest baseline $b_{\rm{min}}$ and the input pupil
%diameter $d_{\rm{input}}$ i.e. 
%$\gamma_{\rm{max}} = \frac{b_{\rm{min}}}{\gamma_{\rm{max}}}$. The
%output pupil diameter will define the diffracting envelope in the
%image while the longest baseline will define the highest Point-Spread
%Function (PSF) size.

%\begin{figure}[th]
% \centering
%\includegraphics[width=0.5\textwidth]{./Figures/figure_densification2.pdf}
%  \caption{Comparision between a diluted (top) and densified (bottom)
 %   array of pupils. Left column shows the pupil distribution and
%    right the corresponding diffraction figure.}
%  \label{fig:densification}
%\end{figure}

Lardiere et al. 2007\cite{lardiere2007} and Patru et al. 2007 \cite{patru2007} have explored how well
different array architectures, i.e. input pupil distributions, can be densified.
Their conclusions show that the choice of
pupil distribution will lead to different field of views, angular
resolution and halo level and therefore require a compromise based on
the scientific goals. Additionally, it should be noted that projection
effects caused by the source trajectory on the sky were not considered
there, but could in principle be implemented.

%\begin{figure}[th]
%  \centering
% \includegraphics[width=0.9\textwidth]{./Figures/figure_fiberDensification.pdf}
%  \caption{Scheme of the implementation of the direct-imaging interferometric set-up by means of optical fibers \cite{patru2008}. \textit{To JPB: can you please make a new cartoon? 
% We can't use figures from published papers without asking permission.}}
%  \label{fig:fiber}
%\end{figure}

Guided optics (whether fibers or 3D integrated optics) can be of great use
to implement the densification of the input pupil since they simplify considerably the routing of
light. Direct imaging and densification with fibers have already
been explored experimentally by Patru et al. 2008 \cite{patru2008}. 
The scheme is easily scalable to an arbitrary large array of telescopes provided an accurate fringe tracking system for the whole array is available. 

The key limitation of a direct imaging instrument is signal-to-noise in the case of a sparse (e.g. non-redundant) pupil. The fraction of the light in the central core is proportional to $
(N_{\rm t} D^2 / B_{\rm max}^2) \times (\gamma_D/\gamma_B)^2$Ó, where $D$ is the telescope diameter, $B_{\rm max}$ the maximum baseline and $\gamma_D$ and $\gamma_B
$ the magnification factors for telescope pupils and baselines respectively. For example, for a 9-telescope non-redundant array (Golay 1971\cite{Golay:1971}), 
this fraction is 0.36, or 0.326 once the geometric factor of circular pupils fitting in a hexagonal grid is taken into account. For background-limited direct imaging or imaging of faint 
structures around a bright star, the signal-to-noise is reduced by this factor. Recovering this signal-to-noise is possible by analysing the direct image with methods analogous to 
aperture-mask interferometry, however those techniques require $\lambda/\Delta\lambda > (B_{\rm max}/D) \times (\gamma_B/\gamma_D)$. For the fields of view required for PFI, this 
either means another significant sensitivity loss with small $\Delta \lambda$, or an integral field unit, removing the simplicity of the direct imaging scheme and making the beam 
combiner similar to any other all-in-one combiner. 

The strength of the direct imaging combiner architecture is in a redundant configuration, where the densified pupil is fully filled. This requires a much greater number of telescopes, 
where $N_{\rm t} \approx (\theta_{\rm FOV}/\Delta \theta_{\rm})^2$. This configuration would only be competitive if relatively small telescopes were very much more affordable than 
large ones\cite{Ireland:2016}.

\subsection{Fibered beam combiners}
The advantage of using optical fibers to perform efficient modal filtering and improve the precision of interferometric measurements by simultaneous photometric correction has been 
demonstrated long ago \cite{CoudeDuForesto:1997}. Two schemes of combination rely on the properties of fibered photonic components: 1) pairwise combiners based on fibre 
couplers \cite{FLUOR}, and 2) all-in-one multi-axial combiners \cite{MIRC}.
Simple fibered 2x2 couplers operating in the H and K bands were used in pioneering photonic beam combiners such as FLUOR \cite{FLUOR}. Experiments with L-band couplers 
were carried out \cite{Mennesson:1999}, but have never been used in an instrument delivering science data.
The pairwise combination scheme could be easily scaled up to multi-telescope arrays by means of a cascade of fiber couplers needed to distribute and combine 
simultaneously the starlight on all possible baselines. However, the main limitations of this approach are the problematic control of differential dispersion and the 
sensitivity of fibers to thermo-mechanical noise, which both grow with the total length of the fibre patches.

\begin{figure}[t]
  \centering
  \includegraphics[width=0.45\textwidth]{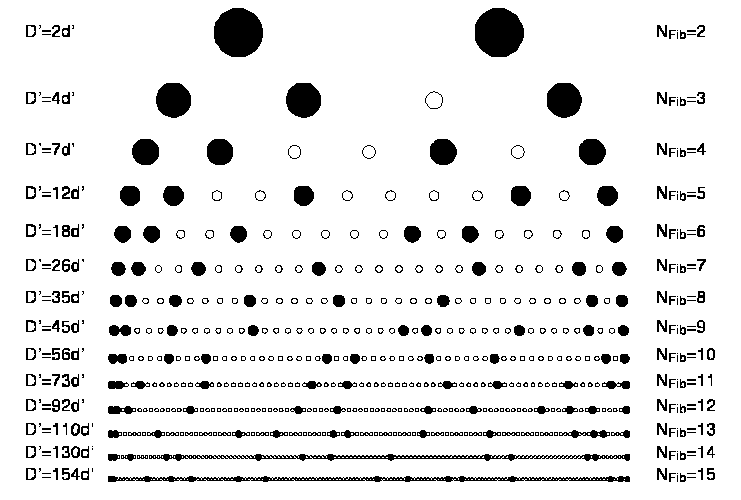}
  \quad
  \includegraphics[width=0.45\textwidth]{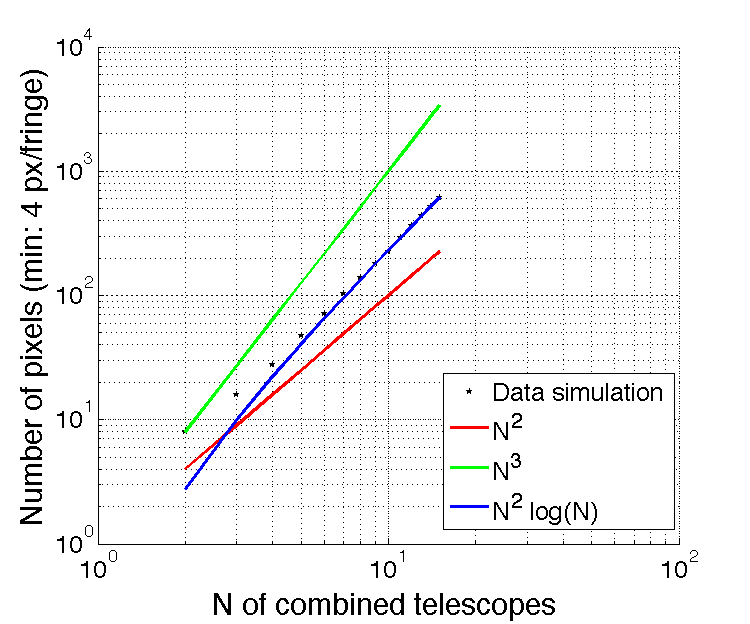}
  \caption{Left: non-redundant fiber arrays optimized for compact beam combination of up to 15 telescopes (from Lacour 2010\cite{Lacour2010}). Right: number of pixels per 
  wavelength channel needed to code all the information of the baselines, as a function of the combined telescopes (stars). A minimal sampling of 4 pixels per fringe is assumed. 
  Red line: $N_{\rm t}^2$ scaling law. Green line: $N_{\rm t}^3$ scaling law. Blue line: $N_{\rm t}^2\log N_{\rm t}$ scaling law.}
  \label{fig:nonredundant}
\end{figure}

All-in-one, multi-axial, fibered beam combiners (see e.g. MIRC\cite{MIRC} and FIRST\cite{FIRST}) are in construction identical to their bulk-optics analogues, but use single 
mode optical fibers for modal filtering of the PSF of the telescopes and deliver the collected light to a V-groove, where the fiber-ends are arranged to form a non-redundant 
linear array.
To date, up to 15 channels were successfully combined in the laboratory \cite{Mozurkewich:2010}, and up to 9 were used on-sky in 
prototype instruments \cite{FIRST}. A non redundant configuration for 24 telescopes has been already calculated\footnote{D. Mozurkewich, personal communication.}.
As mentioned before, the physical length of the non-redundant array grows steeply as $N_{\rm t}^2\log N_{\rm t}$ (see Fig. \ref{fig:nonredundant}). 
Although the use of fibers and micro-optics can reduce considerably the size of the multi-axial 
combination scheme as compared to a bulk-optics design, a cryo-vacuum vessel of relatively large footprint is still needed to host the beam focusing optics. 

A possible amelioration of the scheme is the arrangement of the fibers in a 2D non-redundant array\cite{Lacour2010}. This would result in a more compact optical arrangement, 
however the main difficultly would be the implementation of a cross-dispersion scheme, \textit{e.g.} by means of a highly-spatially-resolved integral field unit sampling the fringes.
A general limitation of fibered beam combiners is related to the long term phase drifts of thermo-mechanical origin, which can only be partially removed by closure phase 
techniques.
A possible solution to this problem could be provided by rescaled versions of integrated optics pupil remappers, such as the Dragonfly instrument\cite{Jovanovic:2012}. 
As discussed in the next paragraph, the 
resilience of integrated optical components to thermo-mechanical perturbations makes them inherently phase-stable. The technological challenge would mainly be related to cope 
with the dimension constraints imposed by the need of coupling many large-diameter, free-space input beams into the reformatting waveguides. 

\subsection{Integrated optics beam combiners}
\begin{figure}[b]
\centering
\includegraphics[width=1\textwidth]{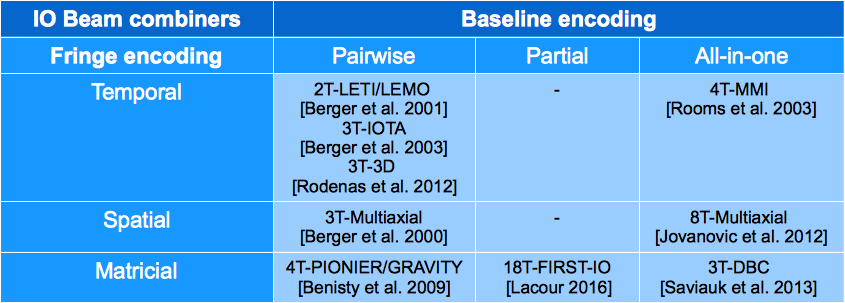}
\caption{\label{tab:IOBC} Existing (TRL$\ge4$) IO beam combiners classified according to Le Bouquin et al. 2004\cite{LeBouquin:2004}}
\vspace{0.0cm}
\end{figure}

First proposed nearly 20 years ago by Malbet et al.\cite{Malbet:1999}, integrated optics (IO) beam combiners represent an effective and compact way to combine interferometrically 
the light collected by many telescopes. IO beam combiners leverage on the modal filtering properties of fibers/waveguides and an enhanced thermo-mechanical stability, due to 
the inherent rigidity of the IO component substrate. Miniaturization and absence of alignment degrees of freedom add up to the advantages of this technology.
To date, several IO beam combiners have been proposed and/or tested in the laboratory and on-sky.  
Fig. \ref{tab:IOBC} is an attempt to sort the various developed IO components according to the conventional classification scheme of the beam combiners outlined in the introduction.

\vspace{-0.2cm}

\begin{figure}[b]
\includegraphics[width=0.40 \textwidth]{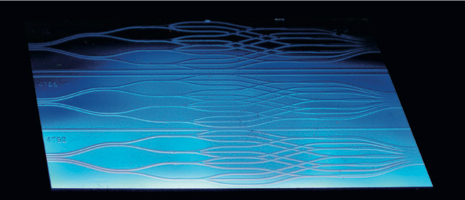} \quad
\includegraphics[width=0.27\textwidth]{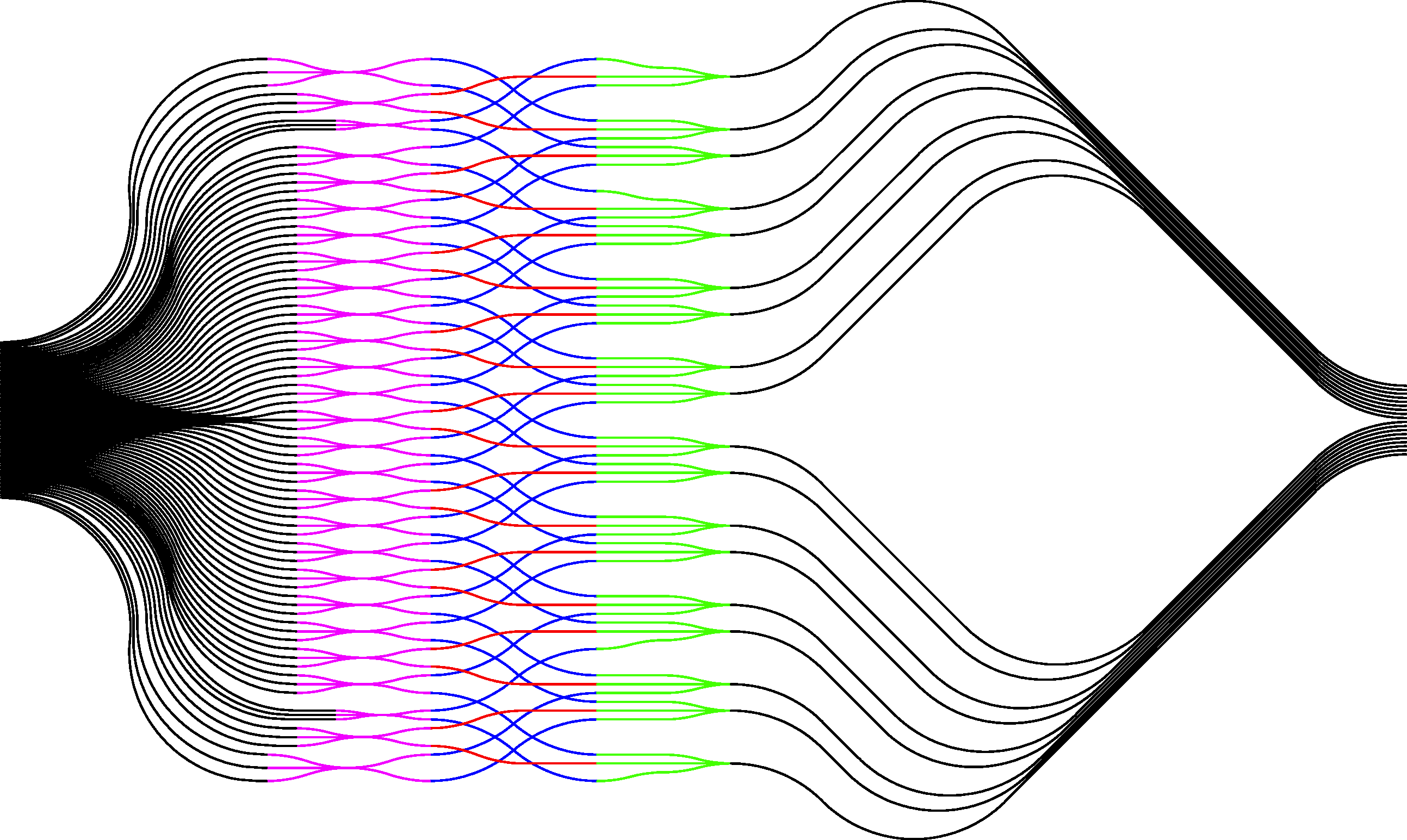}
  \quad
  \includegraphics[width=0.23\textwidth]{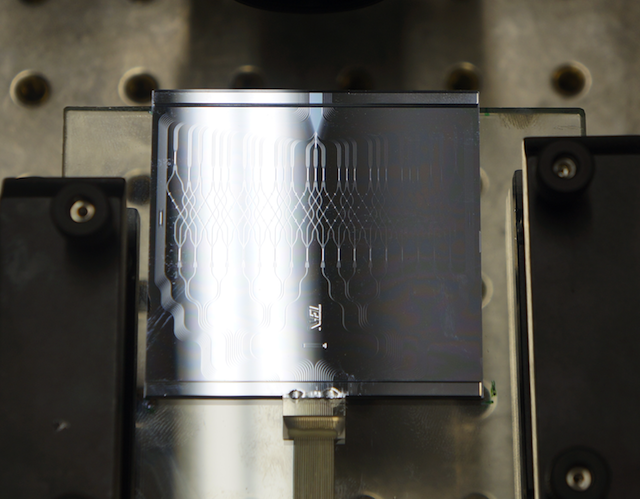}  
\caption{\label{fig:2DIOBC}Examples of planar IO beam combiners. Left: Planar integrated ABCD beam combiner used in the instrument PIONIER at the VLTI. Center: schematic of a 18T ABC beam combiner. Each beam is split into 3 beams, and 32 baselines are combined. The number of output channels corresponds to 96 beams. Right: realization of the 18T beam combiner. The schematic is similar, except for 8 additional photometric channels, giving a total of 104 outputs.}
\end{figure}

Pairwise planar IO beam combiners are the most developed devices so far, which have already been successfully tested on sky to combine simultaneously up to 4 telescopes 
both in temporal or matricial mode fringe encoding \cite{Berger:2001, Berger:2003, LeBouquin:2011}. Additionally, a laboratory demonstration of a pairwise/spatial beam 
combiners was reported in Berger et al. 2000\cite{Berger:2000}. 
These components were operating in H- or K-band, leveraging on the high maturity of ion-indiffusion or silica-on-insulator planar IO technologies, 
which have been developed for telecom applications. 
Silica-on-insulator technology has been also used for manufacturing the partial/matricial beam combiner developed at the Observatoire de Paris Muedon, which combines 
simultaneously 18 channels over 32 selected baselines for the FIRST-IO instrument\footnote{S. Lacour, personal communication} (see Figure \ref{fig:2DIOBC}).
From the PFI perspective, conventional pairwise, planar-IO beam combiners measuring all baselines have limitations regarding the scalability to 1) a large number of telescopes, and 2) to mid-infrared wavelengths.   

The scalability of planar IO pairwise beam combiners to a large number of telescopes is not easy mainly because of the 2D design constraints. 
To combine all baselines simultaneously, a large number of waveguide cross-overs are necessary, which could increase the channel cross-talk  and reduces the SNR of the 
measured visibility, as compared to the performance of the ideal pairwise combiner. Even though the GRAVITY 4T beam combiner achieves cross-talks well below the 1\% 
level, keeping the cross-talk low becomes more and more challenging as the number of combined telescopes increases. 
Moreover, planar IO pairwise combiners  for mid-infrared wavelengths are  still at a research development level as only very recently low propagation losses 
($\sim$0.3 dB/cm) were possible in waveguides manufactured by photolithographic techniques in chalcogenide materials.

In this respect, a promising avenue has been indicated by Rodenas et al. 2012\cite{Rodenas:2012}, where the manufacturing of 3D-IO, 3-telescopes pairwise, mid-infrared beam 
combiner by means of  Ultrafast Laser Inscription (ULI \cite{Nolte2004,Thomson2009}) in a chalcogenide glass was reported. ULI could potentially solve both the problem of cross-
overs (resorting to 3D collision avoidance) and the scalability to longer wavelengths offering low propagation losses in mid-infrared-transparent materials (see also Section 4). 
Additionally, the manufacturing costs are much lower than conventional photolithographic methods.

With the exception of the already mentioned Dragonfly pupil remappper for multi-axial combination\cite{Jovanovic:2012}, existing all-in-one IO combination schemes have 
mostly been tested in the laboratory.  Co-axial mixing of all input fields at the outputs of an integrated optical chip can be accomplished by means of a multimode interference coupler 
(MMI \cite{Rooms:2003}) or by a periodic 2D array of evanescently coupled waveguides (the so called Discrete Beam Combiner - DBC\cite{Minardi:2010}). 
The operation principle of both devices is similar, as multi-field interferograms are obtained at the output of the IO device from the interference of higher-order waveguide modes 
(MMI) or waveguide array supermodes (DBC). So far, MMI have been designed only in temporal scanning mode\cite{Rooms:2003}, but the matricial mode is also possible for MMI 
devices featuring at least 
$N_{\rm t}\times N_{\rm t}$ outputs \cite{LeBouquin:2004}. DBC have been tested in the lab both in matricial \cite{Minardi:2012} and temporal scanning mode \cite{Saviauk:2013}.  
An interesting feature of MMI and DBC devices is that these devices can be designed to be intrinsically short and without waveguide bends, thus limiting the losses of the 
component to a bare minimum. A main drawback of these approaches is the strong chromatic response of the devices. However, experiments have shown that calibrated DBC 
operated in low-resolution spectro-interferometric mode ($R=50$) can be used over a typical astronomical bandwidth with fairly uniform performance\cite{Saviauk:2013} (SNR 
within 20\% of best value across R-band).   
Scaling of DBC devices beyond 8 telescopes\cite{Errmann:2016} is currently limited by the absence of an algorithm predicting the optimal design parameters of the combiner,  
forcing the use of direct numerical optimization algorithms, which scale factorially with the number of combined telescopes. However, these type of combiners could be used in 
partial combination schemes, as proposed in the DOMAC instrumental concept\cite{Minardi:2012c}.

\section{SENSITIVITY COMPARISON OF IDEAL MULTI-TELESCOPE COMBINERS}
\label{sec:sections}
For PFI, the use of a sensitivity optimized beam combination architecture is of foremost importance to warrant high imaging performance at minimal infrastructure cost.
In fact, the a major cost driver for large interferometric arrays are the telescopes, whose unit price notoriously scales harshly with the diameter\cite{Ireland:2016}, 
The overall performance of a beam combination scheme depends on three key elements, namely the intrinsic (ideal) sensitivity of its architecture, 2) the numerical stability of the 
coherence retrieval algorithm, and 3) the fabrication imperfections. The intrinsic sensitivity of the architecture can be roughly evaluated from the signal-to-noise ratio ($SNR$ - defined 
as mean value over standard deviation) of a visibility measurement which, in the absence of thermal background, can be evaluated as\cite{Glindemann}:
\begin{equation}
 SNR=|V|\cdot\sqrt{2\cdot T_{\rm out}\cdot I},
 \end{equation}
where $|V|$ is the visibility modulus, $T_{\rm out}$ is the overall throughput of an individual beam combiner output, and $I$ is the number of photons collected by a single telescope 
during the interferometric measurement. The numerical stability of the coherence retrieval algorithm is crucial to avoid large errors in the determination of the complex visibilities 
from noisy interference data.
While these two factors can be evaluated theoretically from basic principles, the impact of fabrication imperfections can be assessed only with comparative 
experimental tests on existing prototypes.   
Aim of this section is to derive the intrinsic sensitivity on the basis of a simple numerical model describing the three types of beam-combiner architectures, namely the non-
redundant multi-axial array, the pairwise ABCD, and the DBC. 
While this analysis cannot give a definitive assessment of the performance of a realistic beam combiner, it nevertheless helps setting global loss margins for which a given 
architecture can perform better than another one.  

The numerical model underlying our analysis is based on the ideal V2PM description of the analysed architectures \cite{Tatulli:2007} and a statistical model of the detection 
photon-shot-noise (detector read-out-noise is ignored, the approximation is equivalent to the so called "photon-rich" detection regime). We also ignore the contribution of 
thermal background, as it  strongly depends on contingent parameters of the telescope array and cold optics 
environment (see also Ireland et al. 2016\cite{Ireland:2016}).     
The model is used in a Monte-Carlo simulation to extract expectation values of the $SNR$ of the fringe visibility amplitude for an unresolved target ($V=1$), as a 
function of the detected flux per-telescope (assumed to be exactly equal for all telescopes). In each realisation step, the input field of each telescope is calculated as a constant amplitude multiplied by a random phase term. The input fields are fed to the V2PM 
of the beam combiner to derive the expected detected photon numbers at each of the output pixels of the instrument. A noise realisation with gaussian distribution and amplitude 
equal to the square root of the detected photon number is then added to the signal of each output pixel before retrieving the input coherences by applying the P2VM (Moore-
Penrose pseudo-inverse of the V2PM). The coherences are then used to calculate the visibility amplitude for each baseline of the input array. From a statistics of the retrieved 
visibility amplitudes over all calculated realisations (in our case 1000), we could estimate the $SNR$ of each combiner architecture for a range of detected photons per-telescope. 

\begin{figure}[t]
\centering
\includegraphics[width=\textwidth]{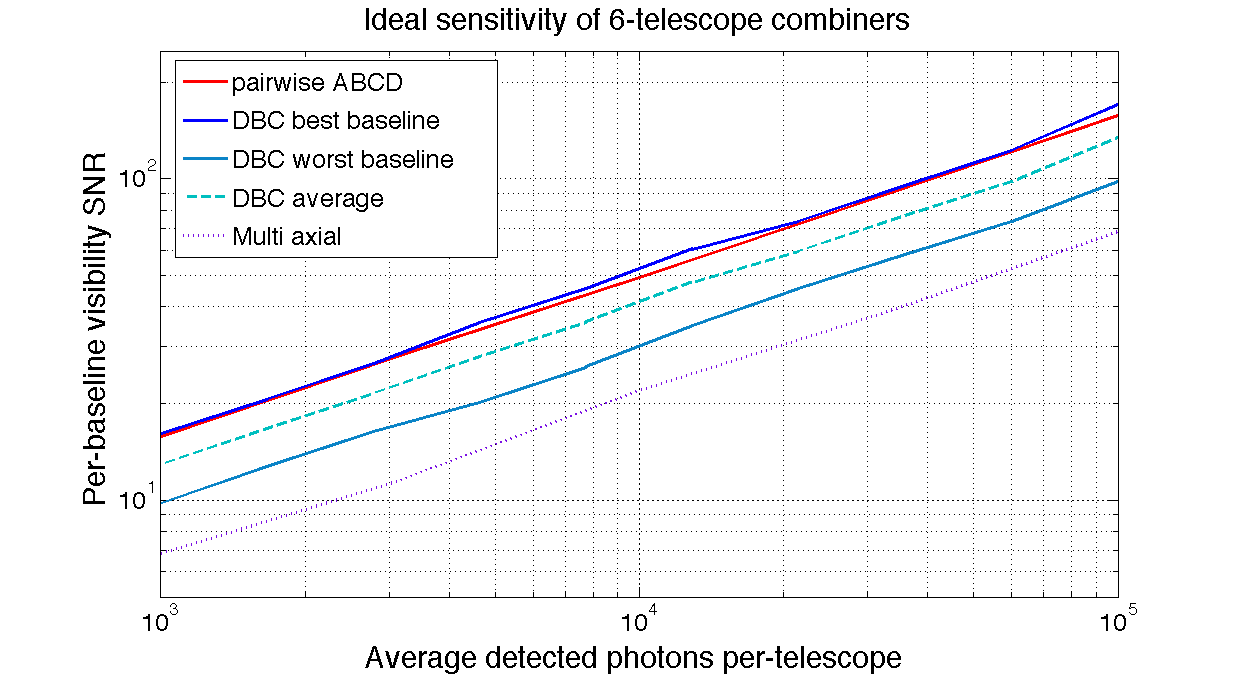}
\caption{\label{fig:simulations} Visibility amplitude SNR as a function to the detected flux per-telescope for ideal 6T ABCD (red line), DBC (blue lines) and multi-axial (purple dashed line) beam combiners. Only the effect of the photon shot noise was take into account in this simulation (see text for details).}
\vspace{0.0cm}
\end{figure}

To allow a fair comparison between the three investigated architectures, we limited our simulations to the case of arrays composed by 6-telescopes, as it enabled the comparison 
of the largest set of combiners. The V2PM of the pairwise ABCD follows design of Benisty et al. 2009 \cite{Benisty2009} scaled up to 6-telescopes and includes 6 1x5 splitters with 
each output  ending up in a 2-way 50/50 splitter, whose ends are connected by 2x2 50/50 directional couplers according to the required baselines combinations. Ideal $\pi/2$ phase 
shifts are assumed to generate 
the ABCD phase pattern, but no cross-over induced cross-talk between channels is modelled.  The DBC design follows the indications of the optimal 6 telescope configuration  
identified for square waveguide arrays in Errmann\&Minardi 2016\cite{Errmann:2016} and which consists in an array of $9\times9$ waveguides.
The multi-axial combiner was modelled assuming the non-redundant configuration displayed in Fig. \ref{fig:nonredundant}, a 4 pixel sampling for the fringes with the 
highest spatial frequency, and a window encompassing 2 full fringes at the smallest spatial frequency. A 5\% photometric pick-off for each channel and a gaussian envelope of the  
individual beam with a waist 2.5 times smaller than the window were additionally assumed.
Large variations of the design parameters of multi-axial scheme did not change qualitatively this picture ($SNR$ variations $\pm 7$\% depending on design parameters).
All simulations were carried out in the monochromatic approximation assuming that dispersive effects 
are negligible, \textit{e.g.} because of dispersion control in the optical setup or sufficiently large spectral 
dispersion at the output of the device.
Results of the simulations are displayed in Fig. \ref{fig:simulations}. 
A comparison between the plots shows a qualitative difference between the three architectures. The first result is that the ABCD and DBC architectures offer a $\sim 2$ times 
better SNR than the multi-axial architecture for a given detected flux. Due to the scaling law of the SNR to the detected photon flux, this would mean that an ABCD/DBC component 
could be about 5-6 dB more lossy than the multi-axial scheme, and still perform better (loss margin). For combiners featuring the same global losses this is equivalent to a 
sensitivity gain of $\sim1.6$ magnitudes ($\sim 1.2$ magnitudes for the DBC in average) respect to a comparable multi-axial scheme. 
These results do not take into account covariances between visibility amplitudes, which mean that for a large number of baselines the uncertainty in measuring individual telescope photometry may have little influence on imaging signal-to-noise, despite causing correlated errors on individual baselines.

Interestingly, and differently from the other two investigated architectures, the DBC scheme features a baseline-dependent sensitivity. The best DBC baselines offer an 8\% better 
SNR, while the worst ones a 35\% lower $SNR$ compared to the ABCD scheme (the baseline averaged SNR of the DBC is only 20\% lower than ABCD).
This feature may find an application in the optimisation of the telescope array combination configuration for the improvement of the fidelity of interferometric image 
reconstruction. Since our simulations show that the SNR scales linearly with the visibility amplitude of the measured baseline, the beam combiner inputs could be arranged 
to be associated long baselines (which could in general resolve the target) to the most performant combination channel, while the short baselines (for which the target is 
unresolved) would be associated to the less performant combination channel. In fact, this configuration would give as a result a more uniform visibility SNR across the sampled (u,v) 
plane, which could potentially impact the quality of the reconstructed interferometric image.     

\section{TECHNOLOGICAL PLATFORMS FOR MID-INFRARED PHOTONICS}

\label{sec:photolithography}
In this section we review the raw materials and fabrication technologies for fibres and integrated optics suitable for both bulk and photonic mid-infrared beam combination instruments.

\paragraph*{Materials}
The PFI science case will require optical instruments operating over a very broad wavelength range extending from 3 to 30\,$\mu$m. 
This is significantly larger than the near-IR range and in principle more difficult to cover with one single material or technology.  
The availability of transparent materials encompassing the mid-infrared range is  large and diverse. 
Suitable materials are typically expected to have good chemical and mechanical stability, excellent intrinsic transparency, low toxicity, a temperature working range compatible with 
cryogenic operations, relatively low index of refraction to minimise the Fresnel losses, limited birefringence for passive functions, no excessive fragility and low ageing. In some 
cases they need to be suitable for thin film deposition by sputtering or photo-evaporation when lithography or etching process are to be implemented. In most cases, they need to 
be suitable for the manufacturing of rather small cross-section waveguides if single-mode operation is sought.
Depending on the application and spectral range, materials that have {\it already} been used for mid-IR photonics are: 

%MIR range is comparatively very large with respect to NIR
\begin{itemize}

\item Niobate Lithium presents also some interest in particular for high-frequency active functions. However the competitiveness of this glass in terms of transparency for wavelengths larger than 2.5 microns does not appear very promising.

\item 2.5--4\,$\mu$m: Fluoride glasses were thought for long to be the ``gold mine'' for mid-IR photonics and optical fibres with theoretical intrinsic losses down to 0.001 - 0.01dB/
km. In reality losses are closer to $\sim1 $dB/km, mainly because of the presence of numerous scattering centres due to impurities in the fabrication process. Fuoride glasses 
remain very competitive solutions for operation below 4 $\mu$m.  

\item For wavelengths up to $\sim$10\,$\mu$m, chalcogenide glasses have shown to be a solution with constantly increasing reliability and promises\cite{Eggleton:2011}. Initially 
considered as difficult glasses to work with, they have reached commercial maturity and are available under non-toxic form to fabricate for instance mid-IR lenses and fibres (e.g. 
with BD-1 and BD-2 glasses). The transparency is excellent, although comparable issues to fluoride glasses may arise due to the presence of impurities and scattering centres.

\item Covering wavelengths longer than 10 microns can be done in principle with telluride-doped glasses or silver-halide glasses. Experimental evidence has been shown by 
Vigreux et al. 2011\cite{Vigreux:2011}. However the intrinsic transparency seems low.

\item In the 20 to 30 microns range, silicon- and germanium-based "far-infrared" waveguides are proposed on a theoretical basis but have not been, to the best of our knowledge, 
investigated for astronomy (Soref et al. 2006\cite{Soref:2006} and references therein).
\end{itemize}

\begin{figure}[t]
\centering
\includegraphics[width=1\textwidth]{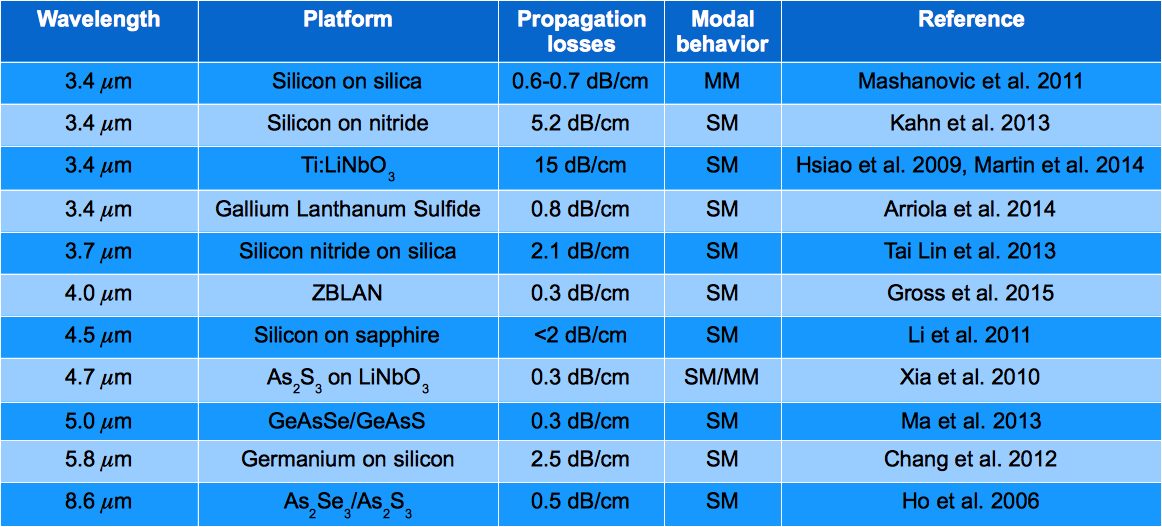}
\caption{\label{tab:IOlosses} Summary of the reported propagation losses of waveguides manufactured with various mid-infared IO technological platforms.}
\vspace{0.0cm}
\end{figure}
\vspace{-0.2cm}

\paragraph*{Fibers}

The availability of mid-infrared fibres is a major aspect for the application of mid-IR photonics to astronomical instrumentation. In particular for stellar interferometry a fibered interface between the telescopes and the recombination unit has proven to be a reliable solution to obtain a workable instrument. In general the use of fibres allows a simplification of the general optical design of the instrument. \\
Significant progress has been achieved in the last years, which led to the industrial maturity of a number of fibre solutions.

\begin{itemize}
\item 2--5 microns  single-mode Zirconium/Indium fluoride fibres with 0.2 dB/m losses are commercially available at a competitive price. In the OHANA experiment\cite{Kotani:2005}, customized fibers with dB/km losses were fabricated. These fibres appear as the prime solution to cover the K, L and M astronomical bands.
\item Chalcogenide commercial fibres up to 9 microns, SM and MM, with different level of dB/m losses. Multimodes up to 9 microns transparency achieve ~0.2 dB/m as well. Very recently, chalcogenide polarisation maintaining fibres with losses below 0.5 dB/m in the mid-infrared here manufactured \cite{Caillaud:2016}.
\item Large-core hollow-core ($\sim1$ mm) multimode fibres for 10 microns light transport commercially available with 0.1 dB/m and zero dispersion.
\item Very low-loss research grade micro-structured fibres with 0.3 to 0.03 dB/m in the L band \cite{Yu:2012}.
\end{itemize}

\paragraph*{Mid-infrared integrated optics}

Planar and 3D IO manufacturing processes suitable for MIR wavelengths are still in a 
development phase, and do not currently offer the quality of silica based micro fabrication. 
In Table \ref{tab:IOlosses} a non-exhaustive list of some of the leading results that have been achieved in the L, M and N bands is presented. Only very recently a few 
technology platforms achieved sub-dB/cm propagation losses in the mid-IR. But low propagation 
losses are only one of the requirements for a useful beam combination technology. It will, for example, be essential to realise integrated components that utilise low cross-talk 
waveguide crossings and high quality beam splitters that are ideally achromatic, and that are suitable for efficient coupling to mid-IR transmitting fibres. 
Here we highlight a few technologies which have already been considered for applications to stellar interferometry: 
\begin{itemize}
\item Deposition of thin films followed by chemical etching or lithography. This is a direct extension of processes successfully used at shorter wavelengths. Single- 
and multi-mode channel waveguides have been demonstrated \cite{Xia:2010, Ma:2013}. 
\item (CW- and ultrafast) laser writing is a recent but promising technique relying on the use of ultrashort laser pulses to directly inscribe three-dimensional waveguide structures 
inside a transparent glass substrate. Ultrafast laser writing provides the ability to fabricate waveguide crossings with zero cross-talk and beam combiners based on two-
dimensional arrays of coupled waveguides (DBC, see previous section). Two- and three-telescope structures have been successfully realised for mid-infrared (Labadie et al. 
2011\cite{Labadie:2011}, \cite{Rodenas:2012}; Arriola et al. 2014\cite{Arriola:2014}, Gross et al. 2015\cite{Gross:2015}, Tepper et al. 2016\cite{Tepper:2016}). Low propagation 
losses in the order of 0.2 dB/cm have also been reported in ZBLAN\cite{Gross:2015} and GLS\footnote{R.R. Thomson, personal communication}.  A current limitation of the 
method is the generation of a remarkable, long-range stress-birefringence (see Diener et al. 2016\cite{Diener:2016}), which could constrain the realisation of complex 3D 
photonic structures. 

\item Ion-exchange and diffusion in glasses. This platform has been very successful in the telecommunication range and is at use for the visible in astronomy. It has been applied at 
longer wavelength in Germanate glasses. Channel waveguides for the mid-IR have been fabricated but only tested at 1.5 $\mu$m \cite{Grelin:2008}. 2-telescope and 3-telescope 
Ti:diffusion Niobate Lithium combiners for the L band have been also fabricated with this platform \cite{Hsiao:2009,Martin:2014}.
\end{itemize}

% Alternative to MIR materials is air-propagation, essentially through fibers with 'hollow core'.  Possibility to evacuate some large core hollow fibers. 
% no material/wet air dispersion.

\section{TECHNOLOGICAL CHALLENGES}

Our review of beam combination schemes and fabrication technologies lead us to identify several technological challenges which need to be overcome in order to enable 
efficient mid-infrared, multi-telescope beam combination schemes suitable for PFI. 
We can distinguish two main challenges, which relate to the identification of 1) the beam combination 
architecture and, 2) the technological approach.

The first challenge is to select the best beam combination architecture for a given interferometric array. 
The solution to this problem include both array architecture and image 
reconstruction considerations and couples to the choice of a suitable technological platform and beam 
combination scheme for science and fringe tracking.
From the respective of the beam combination instrument, the main choice is between a scheme 
allowing the combination of all possible baselines or a sub-set of them 
(\textit{e.g.} for efficient fringe tracking).  
The imaging capabilities of the array could depend from a trade-off between the number of baselines 
and the $SNR$ of visibility measurements 
on individual baselines. The combination of a sub-set of baselines could increase the $SNR$ at the 
expense of (u,v) plane coverage, thus a compromise has to be found.   
This choice should be also confronted with the technical feasibility of an instrument suitable for the 
combination of all baselines. As we have seen in Section 2, the scalability of 
existing beam combination schemes up to $N_{\rm t}=21$ telescopes is not trivial, due to the rapidly 
increasing complexity of fabrication constraints.

A further challenge is to understand which technological approach (bulk optics or photonic) could 
deliver the best interferometric performance in terms of sensitivity and precision.
Given that the PFI design assumes adaptive optics correction for all telescopes, modal filtering may not 
be required anymore for the achievement of high precision visibility measurements, as 
long as high order aberration correction is available. 

In this respect, bulk optics combiners could exhibit a better throughput than photonic ones, as 
the coupling efficiency into a single mode waveguide of the PSF of an aberration-free telescope with 
circular pupil cannot be greater than 80\% \cite{Shaklan:1987}. However, very recent advance in 
micro-structured fibres demonstrated that coupling efficiencies of up to 93.7\% could be achieved 
for beams shaped as ideal PSF of a telescope with circular aperture\cite{Gris-Sanchez:2016}.  

\begin{figure}
\includegraphics[width=\textwidth]{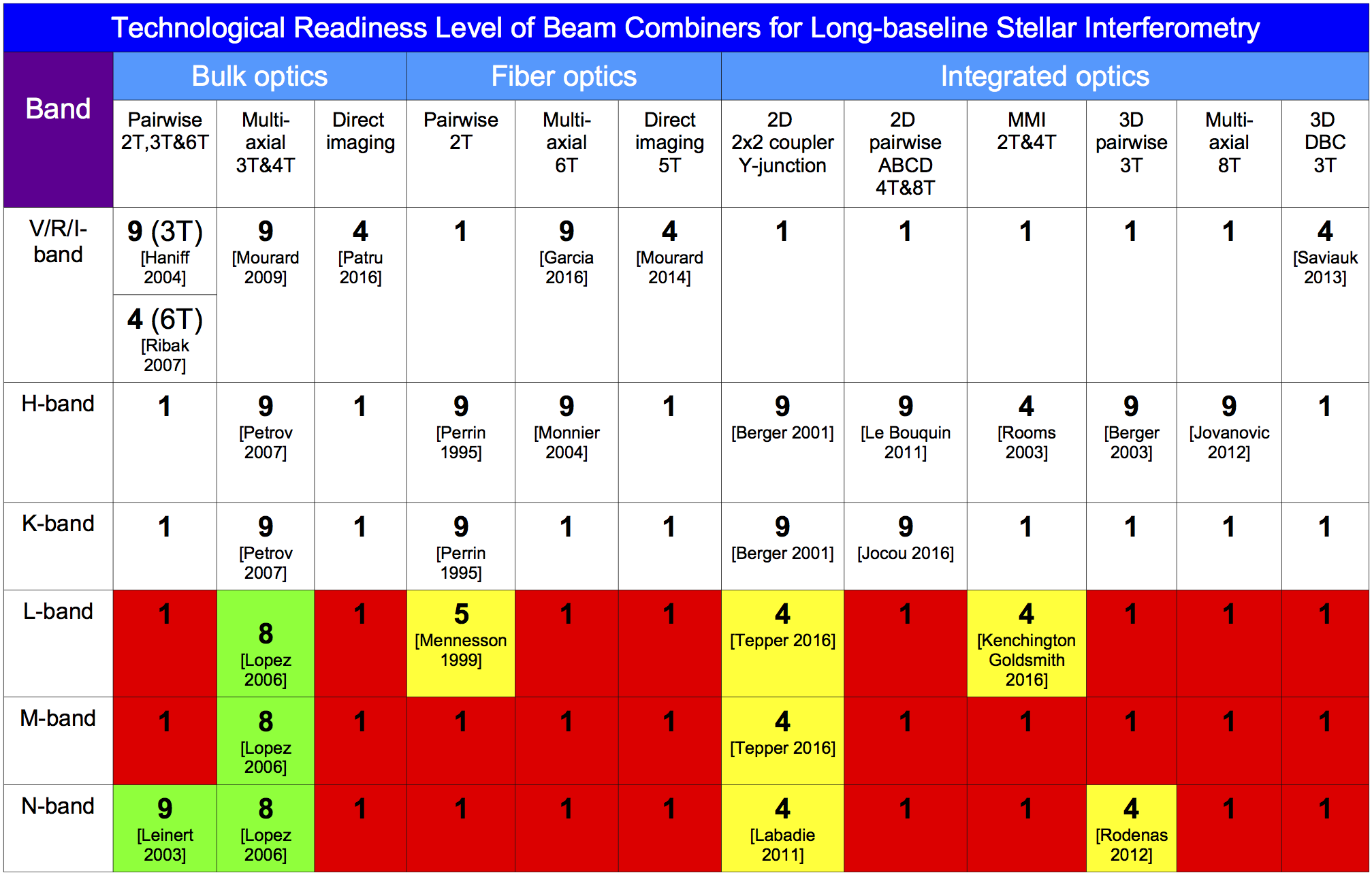}
\caption{\label{tab:BCbands} Resuming table of the TRL of various beam combiner technologies for 
V$^2$ long-baseline stellar interferometry divided by astronomical bands. TRL 1: concept. TRL 4: lab 
test. TRL 5: validation on sky.  TRL 8: instrument qualified. TRL 9: instrument operative.}
\end{figure}

In the short term, a further advantage of bulk optics solutions for the mid-infrared is that this 
technology has already been tested on sky, as indicated in Fig.~\ref{tab:BCbands}, where we attempted 
to the list the Technological Readiness Level (TRL\cite{nasa}) of V$^2$ beam combiners for stellar 
interferometry, sorted by operating astronomical band and technology is shown. 

Even though photonic beam combiners for mid-infrared have reached at most the level of laboratory 
demonstration, integrated optics solutions are in a good position to reach soon the 
level of on-sky test for a moderate number of combined telescopes (e.g. 
$N_{\rm t}=4$), considering the recent successful laboratory test of building blocks like 2x2 couplers
\cite{Gross:2015,Tepper:2016}, pairwise 3-telescope combiners\cite{Rodenas:2012} and the reported 
on-sky test of a two apertures mid-infrared nuller\cite{Norris:2016}.

Nonetheless, current critical points of photonic combiners are the propagation losses and the 
polarisation state control, which require further R\&D . 

%Regarding IO solutions, we point out that the main challenge is to manufacture a component with high 
%throughput (overall losses $<$3 dB).
%As we have seen, only very recently mid-infrared photolithographic IO technologies demonstrated 
%propagation losses below 0.5 dB/cm, which is still high compared to what 
%is possible at shorter wavelengths.
In this context, the recent achievement of mid-infrared waveguides with propagation losses in the order 
of  0.2 dB/cm makes ultrafast laser inscription a very promising technological 
platform for the manufacturing of mid-infrared IO components, offering also very low production costs 
as compared to photolithographic processes. 
A current limitation of the ULI technique is however the low achievable index contrast (from $
\sim10^{-4}$ to $\sim10^{-3}$ depending on the substrate material), which imposes 
large curvature radii for bended waveguides in order to suppress radiation losses. This translates in 
very elongated IO components which could result in beam combiner lengths 
of the order of several centimetres, implying both a higher impact of intrinsic losses and more difficult 
control of the optical path difference in the manufactured component. 
Additionally, the long range ($\sim100-200 \mu$m) tensile stresses induced in the substrate by the ULI 
\cite{Diener:2016}, could make the control of the transverse uniformity and 
birefringence of complex waveguide circuits rather tricky.

\section{CONCLUSIONS AND RECOMMENDATIONS}
\label{sec:misc}

As outlined in this paper, several competing beam combination and related technological solutions are 
available or will soon be available to meet the requirements of the beam combination instruments at the 
PFI facility, all of them entailing design and/or technological challenges which require further R\&D.

Concerning the fringe tracker beam combination scheme, near-infrared fringe measurement on a 
selection of short/intermediate baselines is required. Some level of 
redundancy in the combination scheme will be necessary to limit the impact of failure of fringe tracking 
on individual baselines. As for the fringe tracker instrument, we believe that 
near-infrared IO will play a central role in delivering a stable and compact device with high throughput. 
As mentioned before, planar IO in near-infrared is already technologically 
mature to deliver such baselines-selective beam combiners. Additionally, ultrafast laser inscription could 
be used to avoid cross-overs in the beam combination scheme and 
thus limit the cross-talk between the combined channels.

Design challenges for the science combiner are mainly related to the identification of the best 
combination architecture (partial or full baseline coverage) compromising between 
$SNR$ and (u,v) plane coverage which could enable high-fidelity and high-dynamic-range images for 
the generic PFI target. This first decision will influence the final choice of the type of beam combination 
instrument which will be eventually adopted by PFI. 
Additionally, the design of a compact and mostly maintenance-free cold optics setup is also an issue to 
reduce the financial investment and operation costs of the instrument.

As discussed on Section 3, fundamental properties of the beam combiner architecture and related 
coherence retrieval algorithm make pairwise ABCD and all-in-one DBC schemes 
intrinsically more sensitive than a multi-axial all-in-one solution by about 1.5 magnitudes. Because the 
costs of the telescopes is the main driver of the overall PFI facility
\cite{Ireland:2016}, essentially more sensitive schemes are in principle best candidates for the beam 
combination instrument.  
We warn however that technological/manufacturing issues like \textit{e.g.} the overall beam combiner 
transmission, scalability to large arrays of 
telescopes, conditioning of the V2PM, polarisation and intrinsic chromatic dispersion of the beam 
combiners can have a more determinant role in setting the sensitivity of the final 
beam combination instrument.    
In this context, the direction of future developments of photonic technologies for the mid-infrared will be 
also a crucial ingredient for the choice of a suitable beam combination scheme and relative 
manufacturing technology. 

As from the current stand of the technological development, we mention that a full-baseline coverage 
for a 21 telescopes array would be a very challenging task for any of the 
discussed beam combination schemes or technologies.  
Partial beam combination schemes by means of replication or temporal multiplexing of intermediate 
size beam combiners (\textit{e.g.} for $N_{\rm t}=6$) could represent  a practical solution, at least in the short term.

\acknowledgments % equivalent to \section*{ACKNOWLEDGMENTS}       
 
S.M. and L.L. acknowledge the financial support of the German Ministry of Research and Education (BMBF) under grant "ALSI - Advanced Laser-writing for Stellar Interferometry".   
(05A14SJA).

% References
\bibliography{PFI_combiners_SPIEv2} % bibliography data in report.bib
\bibliographystyle{spiebib} % makes bibtex use spiebib.bst

\end{document}